# The ICT predicament of new ICT-enabled service

*Wen-Jan Chang[a], Rua-HuanTsaih [b], David C. Yen*[c], Tzu-Shian Han[d]*

[a] Department of MIS
National Chengchi University
No. 64, Sec. 2, ZhiNan Rd., Taipei 11605, Taiwan (R.O.C)
wenjan@gmail.com

[b] Department of MIS
National Chengchi University
No. 64, Sec. 2, ZhiNan Rd., Taipei 11605, Taiwan (R.O.C)
tsaih@mis.nccu.edu.tw

*[c] School of Economics and Business
SUNY College at Oneonta
Oneonta, NY 13820
607-436-3458 (office)
607-436-2543 (fax)
David.Yen@oneonta.edu

[d] Department of Business Administration
National Chengchi University
No. 64, Sec. 2, ZhiNan Rd., Taipei 11605, Taiwan (R.O.C)
than@nccu.edu.tw

# The ICT predicament of new ICT-enabled service

**Abstract**

The advancement of information and communication technologies (ICT) has triggered many ICT-enabled services. Regarding this service, the complementary ICT system involves with customers' devices, industry-wide ICT development and nation-wide ICT infrastructure, which are difficult for any individual organization to control. The ICT predicament is the phenomenon that the complementary ICT system is inferior in delivering the promised service quality of new ICT-enabled service. With the ICT predicament, companies face the decision-making dilemma in launching the new service or postponing the launch. This study proposes a process to resolve the decision-making dilemma regarding the ICT predicament.

Keywords: ICT predicament, Radical innovation, ICT-enabled service, complementary ICT system, iPalace channel.

# 1. Introduction

The applications of information and communication technologies (ICT) are widely and drastically accepted in today's environment. Thanks to the advancement and convergence of information/computing technologies, such as Internet, cable TV, customer databases, free-access user-friendly browser, portable devices (i.e., tablet PC and smart phone), multi-media, and cloud computing. These aforementioned ICT applications have made people, organizations, systems and various heterogeneous devices possible to be linked with each other in a more efficient and economical manner than before. These ICT advancements have further stimulated the promising emergence of ICT-enabled services and greatly changed people's way of living, learning, and doing business. In this study, the ICT-enabled services stand for the service driven by and are tightly associated with the applications of ICT.[1] For instance, there are a number of ICT-enabled services provided via mobile apps (Wikipedia, 2013), IPTV (Jang and Noh, 2011), and websites that utilize mainly video clips to provide a virtual environment experience (Kim and Mattila, 2011). Even though the text-based web service, a predecessor of ICT-enabled service, is still an important channel for handling information query and transaction (Parasuraman et al., 2005), organizations and businesses are eager to take the competitive advantages resulted from launching new ICT-enabled services because of the better service quality.

There are two kinds of new ICT-enabled services and they are (1) the radical innovations that are offerings which may not be previously available to customers or provide new delivery systems for the existing services, and (2) the incremental innovations that are changes made to the existing services which are valued as improvements. In general, there are several challenges existed for service innovation including (1) the capability to protect intellectual and property technologies, (2) the incremental nature of innovation, (3) the degree of integration required, and (4) the ability to build prototypes or conduct tests in a controlled environment (Fitzsimmons and Fitzsimmons, 2004). For instance, from the customers' perspective, a radical innovation in ICT-enabled service (RIIS) may in one hand uplift the perceived usefulness while in the other hand curb the perceived ease of use. Both of perceived usefulness and ease of use are the antecedents of actual use (Davis, 1989); however, they may contradict with each other under a RIIS scenario. In other words, a RIIS may enhance the perceived usefulness because of its useful innovation and better integration, and may reduce the perceived ease-of-use because the complementary ICT system is inferior to providing the desired quality of new service. The complementary ICT system actually involves with the ICT infrastructure and the end devices of customers, both of which are provided by other ICT companies/providers. Organizations and businesses usually cannot have the needed authority and thus control over the performance of the complementary ICT system.

---

[1] The ICT-enabled service is the on-line, real-time, interactive service provided through the ICT application. Even though cognizant of the on-going evolution of service and ICT, researchers categorize the following aspects of ICT-enabled services: (1) being information-abundant; (2) being fulfilled through the ICT applications that function intra-organizationally and inter-organizationally; and (3) having more automation, self-service, and intelligence within the service process. Moreover, ICT-enabled services emphasize the ICT empowerment approach that allows the customer to assume a new, and perhaps more independent, role as an active participant in the self-service process that enhances his/her satisfaction of customization, accuracy, convenience, control and so forth.

Similar to the fact that the development of the transportation infrastructure usually lags behind the development of automobile vehicles, the development of information and telecommunication infrastructure lags behind the development of RIIS. For this reason, it is very likely that the complementary ICT system is inferior to providing the desired quality of new service. Therefore, the CEOs may recognize that the complementary ICT system is likely lack of the needed readiness for launching the new service. Under such a scenario, the CEOs contemplate the coin-flip predicament and that is the potential competitive advantage resulted from pioneering a new service or the possible bad service quality resulted from the lack-of-readiness of the complementary ICT system. This predicament is referred to as the ICT predicament of RIIS in this study. Especially when the new service involves a significant investment, the CEOs are eager to resolve the decision-making dilemma of this aforementioned ICT predicament that leads to conflicted consequences. To the best of our knowledge, however, there is a lack of formal study for resolving/handling the decision-making dilemma caused by the ICT predicament of RIIS. This study addresses this academic gap which is one of the important research contributions.

One way to resolve the decision-making dilemma of the ICT predicament is to uncover whether the advantages of RIIS outweigh its disadvantages. Past researches emphasize that some quality factors are more important than others (Zhang and von Dran, 2002). If customers take important factors into account positively and assess minor factors negatively, they may still perceive the whole service in a favorable manner. Consequently, to obtain the clues for resolving the decision-making dilemma of the ICT predicament of RIIS, this study proposes not only to identify the relevant service quality factors, but also to verify the integrated effect of quality attributes. This finding may prove to be another highlight of this study.

In short, this study proposes the following processes for resolving the decision-making dilemma of ICT predicament of RIIS:

(1) An existing service providing a similar purpose as the new service should be identified as the counterpart. The new service may excel in some attributes but the existing counterpart may surpass in some other attributes.

(2) The causal model for measuring the new service and its counterpart should be established accordingly.

(3) The instrument for measuring the users' perceived service quality regarding the new service and its counterpart should then be developed.

(4) Setting up a survey environment to collect data will be needed.

(5) By analyzing the collected survey data, a comparison of performances can be made between the new service and the counterpart.

(6) In case the performance of the new service surpasses its counterpart, the finding that the advantages of the RIIS outweigh its disadvantages can be supported.

Specifically, this study takes the iPalace channel (Tsaih et al., 2012) of the National Palace Museum (NPM) in Taipei as an instance to discuss the ICT predicament of RIIS. The iPalace channel broadcasts videos of NPM's treasures in an online mode. However, the complementary ICT system is likely lack of the readiness for the desired quality of the iPalace channel because of the potential of huge peak-demand to access it worldwide. In short, the senior managers of NPM may face the ICT predicament regarding the iPalace channel. We apply the aforementioned processes to resolve the decision-making dilemma of the ICT predicament of iPalace channel.

The structure of this paper is organized as follows. In Section 2, the relevant information about the museum and the literature review related to ICT-enabled service and web-based service quality are covered. In addition, the iPalace channel is also briefed in this section. In Section 3, we present the application of the proposed processes to resolving the decision-making dilemma of the ICT predicament of iPalace channel. Finally, in Section 4, a discussion of the findings is presented, along with the conclusion and future work.

## 2. Literature review

*2.1 Museum and ICT-enabled service*

Traditionally, museums have been used to refer to locations specifically designed and used for storing and exhibiting historic and natural objects (Lewis, 1992). The function of a museum is basically to conserve, promote, and conduct researches on objects it houses and also to educate public. Nowadays, with the development of society, museums' foci have been shifted from object-oriented (before the 1980s), education-focused (the 1980 – 2000s) to public-centered (after the 2000s) (Chang, 2011). However, in today's society with abundant sources of entertainment and leisure, museums must make more effort to catch and attract consumers' attention (Hume, 2011; Hume and Mills, 2011). The ICT can play such an important role in this regard. For instance, museums can use the video clips instead of text to help illustrate artifacts. Past researches suggest that video-based presentations can deliver more complete, vivid, tele-presenting, and appealing information to viewers than text-based presentations (Hoffman and Novak, 1996). As a result, ICT-enabled services are growingly introduced by the museum to enhance its experience (Knell, 2003; Marty, 1999; Thyne, 2000). For instance, museums can expand their capabilities by offering Internet technologies, improving their core services to include online displays and material presentation, and better managing their inventory (Hume and Mills, 2011). Prior studies indicate that video technologies have received great attention and become a great resource for enhancing museum's activities and learning (Macdonald, 2002; vom Lehn, 2010).

*2.2 Web-based Service Quality*

There are two basic approaches related to web-based service quality research. One type uses SERVQUAL as a basis and the other one creates new categories for rendering web-based service (Kim and Mattila, 2011).

The former type intends to develop a global measurement for web-based service quality. Most of these researches adopt the E-S-QUAL developed by Parasuraman et al. (2005) as a basis to measure web-based service quality. However, consensus is still lacking because the measurement of web-based service quality is dependent on tasks, users, and sectors (Rowley, 2006). Other researches develop different measurements of web-based service quality for different scenarios. Table 1 list some previous researches related to web-based service quality. Among them, all listed models place a focus on analyzing the E-commerce websites. As a result, other kinds of web-based service are largely overlooked (Schaupp, 2010).

**Table 1**

Relevant research on web-based service quality.

| References | Quality Factors |
| --- | --- |
| Liu et al. (2000) | Information and service quality, System use, Playfulness, System design quality. |
| Kaynama and Black (2000) | Content and Purpose, Accessibility, Navigation, Design and Presentation, Responsiveness, Background, Personalization and Customization. |
| Yoo and Donthu (2001) | Ease of use, Aesthetic design, Processing speed, and |
| | Security. |
| Loiacono et al. (2002) | Ease of understanding, Intuitive operations, Informational fit-to-task, Tailored Communications, Trust, Response time, Visual appeal, Innovativeness, Emotional appeal, On-line completeness, Relative Advantage , Consistent image. |
| Vidgen and Barnes (2002) | Usability, Design, Information, Trust and Empathy. |
| Santos (2003) | Ease of use, Appearance, Linkage, Structure and Layout, Content, Reliability, Efficiency, Support, Communication, Security, Incentives. |
| Parasuraman et al. (2005) | Efficiency, Fulfillment, System availability, Privacy, Responsiveness, Compensation, Contact. |

With a further focusing on museum website service quality literature, it is notable that most of them emphasize primarily on the website elements just as other web-based service quality studies do. For example, Campbell and Wells (1996) suggested three factors: "Appeal factor", "Retention factor" and "Revisit factor". Among them, "Retention factor" can be divided into three sub-factors including "Organized format", "Quality content", and "Personal relevance". Santos (1999) defined four main factors and they are "Usability", "Functionality", "Site Reliability" and "Efficiency". Pallas and Economides (2008) introduced MUSEF (Museum's Sites Evaluation Framework), which is composed of six factors: "Content", "Presentation", "Usability", "Interactivity & Feedback", "e-Services" and "Technical". These service quality factors can be classified and grouped into the above seven categories as well. This may be due to the fact that most museums built their websites as an aid to the physical museum. The museum website is merely to advertise or provide the related information of activities and visits, instead of substituting the traditional role and functions for a physical museum. When museum websites' functions are merely employed as an information channel, the measurement of their service quality is not much different with that of an E-commerce website. However, in this study's context, an artifact-displaying museum website can be utilized to offer online relic appreciation service. By doing so, the corresponding measurement should include some museum's traditional functions. Therefore, the measurement of the service quality of artifact-displaying websites should be adapted from that of ordinary websites to include some museums' native functions.

*2.3 National Palace Museum and the iPalace channel*

Like any other museums in the world, the NPM has also experienced the changing processes of employing the advanced ICT in performing exhibition and marketing endeavors. The NPM has rich collections of ancient Chinese artifacts, amounting to approximately 690,000 items. These ancient artifacts provide precious materials and information for visitors to learn more about Chinese history and ancient arts. In response to the fast ICT advancements and the irresistible wave of the service-oriented demands, the NPM attempts to overcome physical limitations by adopting advanced ICT and new media, including interactive gallery exhibits, personal digital guides, cultural games, online virtual museum, and social net. Some ICT-enabled services are auxiliary by nature to merely support the physical visit activities. Not satisfying with the current scope, the NPM is developing more core-services like ICT-enabled services to deliver a direct customer-centered museum experience. For example, the NPM devoted great efforts to create digitalized image databases for its treasures in recent years (National Palace Museum, 2010). These online databases can indeed make the treasured artifacts and educational resources accessible to the whole world via Internet.

In addition, to further explore the value of digital archives, NPM is developing a video-based website to broadcast the video collections of the Chinese relics it conserves. The NPM aspires to a video-based website that allows people worldwide an access to the video collection of the museum's great Chinese relics, for example, watching an online TV program. Furthermore, the NPM would like to attract the younger generation by providing them with a wonderful viewing experience, to further enrich their lives (Tsaih et al., 2012). In addition, the video channel, which can be easily accessed via portable devices (such as tablet PCs and smart phones), acts as a superb marketing tool, by providing video clips related to exhibitions and the museum collection, as well as enhancing an online visiting experience. These objectives have led to the adoption and development of iPalace channel initiative to systematically present all the treasures of NPM.

Tsaih et al. (2012) stated the strategic vision of the iPalace channel as follows. The service delivery system should always deliver fresh and attractive video contents smoothly to provide well-defined NPM online visiting experiences. Being an online medium of the NPM, the iPalace channel needs to not only meet with the high-quality brand recognition of the NPM, but also be able to cope with the huge peak-demand worldwide. Thus, its main operating strategy is to keep the website fresh and attractive for customers' retention and to maintain effectively a load-balance to smoothly display videos. To this end, it requires a continual provision of new videos and a periodical updating of the interface appearance. For effective load-balancing, the iPalace channel technically requires a task-oriented process design, an elastic web service infrastructure, well established peer-to-peer networking, and efficient multi-task and distributed system architecture.

The service concept is to provide customers with a TV program offering a wonderful viewing experience of NPM artifacts, which can also be contributed to part of the visiting experience. One of key ideas regarding this service concept is to provide the video service via an uninterrupted multicasting method such as a web-based TV channel. Another key idea may be to follow the concept of museum displays to organize the contents by themes, instead of the concept of search-in-warehouse from the discipline of information system. The target audience segment is the younger generation who likes watching videos online rather than reading texts.

The prototype of iPalace channel is built on the cloud-computing infrastructure provided by Chunghwa Telecom's Hicloud. In short, from the museum perspective, the iPalace channel is a radical innovation for the following aspects: (1) It is a service like a TV program using the WWW; and (2) It is a service provided via a cloud-computing infrastructure.

From the customers' perspective, the iPalace channel is also radical since no other museum provides a TV program service on the WWW at the current stage.

With the service quality concern, the iPalace channel not only needs to meet with the high-quality brand image of NPM but also be able to cope with the huge peak-demand worldwide. This means the NPM needs to overcome some difficulties in both content and technology aspects. While the customers' acceptance of iPalace channel is growing fast, the complementary ICT system may be becoming more inferior and obsolete in terms of the smoothly delivering videos to customers. The management team of the NPM, though eager to embrace ICT-enabled services, is wondering how to take a tradeoff between the content and technology requisites.

## 3. The proposed processes

*3.1 The selection of counterpart and the causal model*

The goal of these proposed processes is to investigate the pros-and-cons of a RIIS in comparison with a counterpart service to solve the decision-making dilemma of the ICT predicament. Regarding the counterpart of iPalace channel, the current text-based webpages of the NPM for displaying relics online is first chosen. For certain relics, the text-based web pages are used to display images and some text illustrations of them. With a "zoom in" function, webpage visitors can actually view a higher resolution version of an image. On the other hand, acting like a TV channel, the iPalace channel broadcasts video clips of the same relics with a thorough oral explanation. These video clips are in fact produced by a well-known movie director.

Secondly, to make a comparison between the performance of the iPalace channel and the current text-based webpage, a well-established causal model of service quality factors, customer satisfaction, and behavioral intention is developed accordingly. This model suggests that service quality factors are antecedents of customer satisfaction, and customer satisfaction affects behavioral intention consequently (Carlson and O'Cass, 2010; Johns and Howard, 1998; Oliver, 1980). In the website-related literature, customer satisfaction and behavioral intention are also widely used to assess the success and adoption (Davis et al., 1989; DeLone and McLean, 1992; McKinney et al., 2002; Seddon, 1997; Venkatesh et al., 2003; Wixom and Todd, 2005). Besides, museums are interested in learning the relationship existed between customer satisfaction and the behavioral intention including repeated visiting and word of mouth effect (Harrison and Shaw, 2001). Furthermore, word of mouth advocacy has been cited as a key promotional tool for museums and other cultural institutions (DiMaggio, 1986). From the above discussion, the performance indicators of an artifact-displaying museum website should include customer satisfaction and behavioral intention.

*3.2 Measurement and questionnaire development*

Next, each construct needs to be properly set up. Currently, there is no existing foundational support for evaluating the service quality of artifact-displaying museum website. For this reason, this study re-categorizes the factors of Table 1 (appendix section) into seven groups including content, usability, interface, system reliability, communication, security and fulfillment. These seven categories are content, usability, interface, reliability, communication, security and fulfillment. However, this study neglects the factors of communication and security. The NPM is planning an ambitious RIIS which will be able to provide the following functions such as channel service, social networking service, and culture gaming service via the same cloud platform. Therefore, the design of the iPalace channel does not take the communication and security issues into considerations.

In short, Table 2 shows factors and corresponding sub-factors regarding the service quality of artifact-displaying website as per earlier discussion.

**Table 2**
The service quality factors for museum artifact-displaying website.

| Main Factor | Sub-factor |
|---|---|
| CONTENT | INFORMATION QUALITY |
| | TELE-PRESENCE |
| | MEDIA QUALITY |
| USABILITY | INTUITIVE OPERATION |
| | INNOVATIVENESS |
| INTERFACE | VISUAL APPEAL |
| | CONSISTENT IMAGE |
| SYSTEM RELIABILITY | ACCESSIBILITY |
| | RESPONSE TIME |
| | EDUCATION |
| FULFILLMENT | ENTERTAINMENT |
| | PROPAGATION |

Content refers to the design of presentation and layout of factual information on a website and also the organization of the website's content. It would affect whether or not the website is easy to read and be understood by visitors (Loiacono et al., 2002; Santos, 2003). Besides, the provided information should be accurate, relevant and trustable (Liu et al., 2000; Loiacono et al., 2002). However, to better capture the difference between the video-based and text-based artifact displaying websites, tele-presence is included as one of the sub-factors of content in this study. In fact, tele-presence refers to how well the artifacts are presented through the multimedia technique on the website. Good tele-presence means that the users may feel the same as they are at the scene personally and gaze steadily at the screen while browsing the website.

Usability is related to how easy and efficient the website is for visitors to conduct, learn and master to navigate and search (Loiacono et al., 2002; Parasuraman et al., 2005; Santos, 2003; Yoo and Donthu, 2001).

Interface refers to the proper use of multimedia technology (including color, graphics, image, and animations) with the appropriate size of the web pages that are pleasing to the visitors' eyes and to avoid cluttered pages (Loiacono et al., 2002; Santos, 2003; Yoo and Donthu, 2001). In addition, using a creative and differentiating approach on the website (Loiacono et al., 2002) may make visitors feel novel and exciting. What is more important may be to provoke a positive customer's experience and thus reflect the company's image (Loiacono et al., 2002).

Reliability is defined as the capability to correctly and consistently perform the promised service, and the speed of online processing and interactive responsiveness including downloading, search and navigation (Parasuraman et al., 2005; Santos, 2003; Yoo and Donthu, 2001). Besides, the service providers should have sufficient hardware and communications capacity to meet the peak demand (Loiacono et al.,

2002).

Fulfillment refers to meeting the customers' demand for needed information, and allowing visitors to conduct important business functions over the Web (Loiacono et al., 2002). In other words, it means the extent to which the site's promises are fulfilled (Parasuraman et al., 2005). Since an artifact-displaying website can be regarded as an alternative for visitors to appreciate relics, its goals should include some of the physical museum's native functions related to exhibition such as education (Goulding, 2000; Harrison and Shaw, 2004; Rentschler and Gilmore, 2002; Saleh, 2005), propagation (Goulding, 2000; Harrison and Shaw, 2004; Rentschler and Gilmore, 2002) and entertainment (Rentschler and Gilmore, 2002) in assessing the service quality. The native functions of museums to assess the service quality of artifact-displaying websites are included as sub-factors of fulfillment.

Based upon the existing survey instruments of the measurement shown in Table A2 of appendix section, we developed two versions of questionnaires in order to fit the different characteristics of the iPalace channel and the original text-based website. In this way, we can evaluate, compare and analyze two types of artifact-displaying museum websites with the same proposed service quality factors. As shown in Tables A3 and A4 in the appendix section, measurement items of two questionnaires are basically the same except two items related to assessing specific multimedia usage.

Questionnaires actually comprise three sections. The first section assesses the service quality factors. There are 28 scale items for both the text-based version (see Table A3) and the video-based version (see Table A4) questionnaires. The second section is used to evaluate customer satisfaction and behavioral intention. The six-point Likert's scale was used in the first and second sections. The respondents were requested to indicate their levels of agreement of the statements, with a scale that ranged from 1 (do not agree at all) to 6 (definitely agree). The third section collects demographic data. The questionnaires were examined and revised by domain experts and the NPM staffs in order to ensure their logical consistency and diction accuracy.

*3.3 Survey Administration*

The target customer is young people. Therefore, the young people should be the majority of our samples to test their attitude towards this service. As shown in Zickuhr and Madden (2012), the young generation are more easily and willing to explore things online than other generations. We adopt the convenience method to collect data since it is the most economical way to gather samples. We posted the questionnaires on a survey website. Then, we sent the invitation emails containing the hyperlink of the survey website to college students, who are also the younger generation by nature. We also invite them to forward this email to their groups and/or friends, who are mainly composed of young people. Table 3 summarizes the demographic profile of total respondents from these two samples. It shows that the majority of respondents (79.4%) are young people with ages from 17 to 30.

Both questionnaires, for the text-based and iPalace ones, were posted on these aforementioned websites. Before respondents began to fill out the questionnaire, they were asked to watch a specific artifact displayed on either text-based web pages or iPalace channel. Anyone who clicks the hyperlinks to our web questionnaires is able to participate in our survey. The participants decide to fill only one or both of the two questionnaires. About two third of respondents complete both questionnaires. We treat the comparison of the two questionnaires as from the same groups when using different services.

All the questionnaires were collected within two weeks. When the survey was completed, 337 questionnaires of the text-based version and 387 questionnaires of the iPalace version were collected. After data screening, we eliminated 3 incomplete and repeated questionnaires from each version of questionnaire. As a result, the total effective sample size was 334 for the text-based version, and 384 for the iPalace version.

**Table 3**
The demographic profile of total respondents

| Measures | Items | Frequency | Percentage |
|---|---|---|---|
| Gender | Male | 254 | 35.4 |
|  | Female | *464 | *64.6 |
| Age | < 16 (year) | 10 | 1.4 |
|  | 17 – 23 | *348 | *48.5 |
|  | 24 – 30 | 222 | 30.9 |
|  | 31 – 40 | 70 | 9.7 |
|  | 41 – 50 | 43 | 6 |
|  | 51 – 60 | 24 | 3.3 |
|  | > 61 | 1 | 0.1 |
| Average daily browsing time | < 0.5 (hr.) | 30 | 4.2 |
|  | 0.5 – 1 | 45 | 6.3 |
|  | 1 – 2 | 149 | 20.8 |
|  | 2 – 5 | *343 | *47.8 |
|  | > 5 | 151 | 21 |
| Education level | Junior High School (inclusive) | 6 | 0.8 |
|  | High school or vocational | 17 | 2.4 |
|  | University (College) | *450 | *62.7 |
|  | Master | 216 | 30.1 |
| | Doctor | 29 | 4 |
| | Cultural and creative industries | 32 | 4.5 |
| Career | Non-cultural and creative industry | 239 | 33.3 |
| | Student | *447 | *62.3 |

* denotes the relatively highest value.

Table 4 is the reliability table of each service quality factor on these two samples. Every item in each main service factors shows a high reliability.

**Table 4**

The reliability of service quality factors for these two samples.

| Cronbach's Alpha | | |
|---|---|---|
| Factor | Text | Video |
| Content | .805 | .881 |
| Usability | .910 | .881 |
| Interface | .830 | .882 |
| Reliability | .891 | .861 |
| Fulfillment | .876 | .878 |

After the EFA, we adapted the original assessment model into the new measurement as shown in Table 5. The content and usability factors are integrated into one factor. In addition, there are three items (the tele-presence items of content and the innovativeness item of interface) reclassified into the fulfillment factor. The tele-presence items check about whether or not the users gaze steadily at the screen while browsing the website. The innovativeness of interface item asks if the interface is innovative. These three items can be interpreted as the museum inherited function of exhibition that emphasizes on the exhibition layout and the atmosphere. For this reason, it is reasonable to classify them into the fulfillment factor. Here, we rename the group of these three items as the "exhibition." Then, we had conducted CFA, convergent, discriminant and nomological validity tests to check the validity of this adapted measurement. All the results ensure and support its validity.

**Table 5**

The new main factors and sub-factors of the service quality.

| Main Factor | Sub-Factor |
|---|---|
| CONTENT & USABILITY | INFORMATION QUALITY |
| | MEDIA QUALITY |

| | INTUITIVE OPERATION |
|---|---|
| INTERFACE | VISUAL APPEAL |
| | CONSISTENT IMAGE |
| RELIABIITY | ACCESSIBILTY |
| | RESPONSE TIME |
| | EDUCATION |
| FULFILLMENT | ENTERTAINMENT |
| | PROPAGATION |
| | EXHIBITION |

*1.4 Analysis of survey data*

To assess the relative advantages and disadvantages of the text-based website and iPalace channel, we conduct t-tests to compare these two websites on each service quality factor, customer satisfaction, and behavioral intention. Table 6 shows the mean values, standard deviations, and the t-test results. The results reveal that significant differences exist in most of quality attributes, customer satisfaction, and behavioral intention between these two websites.

**Table 6**
The measurements of service quality factors, customer satisfaction and behavioral intention.

| Factor | Text-based | | iPalace | | T | Sig. |
|---|---|---|---|---|---|---|
| | Mean | S.D | Mean | S.D | | |
| FULFILLMENT | 4.2497 | .7690 | 4.7052 | .69873 | -8.257 | *.000 |
| CONTENT & USABILITY | 4.9336 | .62120 | 4.9010 | .68133 | .666 | .506 |
| RELIABIITY | 4.8114 | .79008 | 4.6595 | .84395 | 2.477 | *.013 |
| INTERFACE | 4.6939 | .73753 | 4.9232 | .72147 | -4.204 | *.000 |
| CUSTOMER SATISFACTION | 4.3533 | .94012 | 4.7109 | .84953 | -5.316 | *.000 |
| BEHAVIRAL_INTENSION | 4.0120 | 1.00442 | 4.3841 | .94163 | -5.120 | *.000 |

* $p < 0.05$

From Table 6, there is no significant difference between these two websites with regard to CONTENT & USABILITY. As to FULFILLMENT, the iPalace channel's mean of 4.7052 is significantly lower than the text-based website's mean of 4.2497 (T = -8.257, p = 0.000). Museum artifact-displaying websites, including the iPalace channel, inherit some museum native functions such as education, entertainment, propagation and exhibition. Video has been widely used as teaching aids. We are also used to watching TV and going to see movies as leisure activities. Compared to text and pictures, videos can give us auditory and visual enjoyment. Besides, they are also a good media to promote culture (Johnston and Bloom, 2010). From the above discussion, it is reasonable to observe that the FULFILLMENT of iPalace channel is higher than that of the text-based webpage.

In terms of RELIABIITY, the iPalace channel is significantly lower than the text-based website. The sole significant sub-factor of reliability is response time. Response time here refers to the loading time of web pages or videos. The problem of managing streams of audio and video in a distributed system was always one of the most popular and fundamental problems addressed by research on computer systems (Jeffay et al., 2001). Since the iPalace channel transfers information through videos, it also suffers the severe problem of bandwidth. Therefore, it is reasonable to observe that the iPalace channel has lower RELIABIITY than the text-based webpage.

As per earlier discussion, the iPalace channel is a full metaphorical graphic interface, with fewer texts. Compared to general websites, it is novel to users. The appearance of interface may be more visually pleasing than general websites are. Besides, when information is presented through video, the audience would be less likely interfered by the background. Additionally, the video also gives audiences deeper impression. Therefore, with regard to INTERFACE, it is reasonable to observe that the iPalace channel is higher than the text-based webpage.

From Table 6, the iPalace channel is significantly higher than the text-based website in terms of CUSTOMER SATISFACTION and BEHAVIRAL_INTENSION. Table 7 summarizes the pros and cons of the iPalace channel regarding the quality attributes.

**Table 7**

Pros-and-cons of the iPalace channel regarding the quality attributes.

| Pros | Cons |
| --- | --- |
| Interface | Reliability |
| ● Visual appeal | ● Response time |
| ● Consistent image | |
| Fulfillment | |
| ● Education | |
| ● Entertainment | |
| ● Propagation | |
| ● Exhibition | |

*3.5 The decision making regarding the iPalace channel*

In sum, the iPalace channel has a better performance on the service quality factors of interface and fulfillment. However, it does not perform well on reliability issues. The reliability problems cannot be resolved easily because of the inferior complementary ICT system. Thus the survey result reveals that there is indeed an ICT predicament occurred in the iPalace channel initiative. Consequently, the management's concerns are reasonable. Nevertheless, this inferiority is not so substantial to hurt the ultimate performance of the iPalace channel, i.e. customer satisfaction and behavioral intention. Our empirical results show that the iPalace channel visitors are more satisfied and willing to recommend and reuse the website than the text-based website visitors are. Therefore, this result implies that iPalace channel, as a radical innovation, is ready to be launched even with the existence of current inferior complementary ICT system.

## 4. Concluding remarks

Nowadays, many organizations seek to identify and create new markets by initiating RIIS, such as online gaming and online teaching. However, the CEOs may recognize that the complementary ICT system is possibly lacking of the needed readiness for launching the new ICT-enabled service. This phenomenon can be referred to as the ICT predicament. In an ever changing and competitive environment, instead of controlling the uncontrollable environmental factors or waiting for their improvement, management should develop ways accordingly to 'probe into the future' (Brown and Eisenhardt, 1997). In this study, we propose a process to solve the decision-making dilemma by building a prototype of the new service and testing it in a controlled environment to gauge its potential benefits and associated risks.

In the case of iPalace channel, the ICT predicament is obversed due to the inferior complementary ICT system. However, the experimental result also uncover that visitors are more satisfied with the iPalace channel and hence, more likely to revisit and recommend to use it. Consequently, the advantage of iPalace channel outweighs its disadvantage. This result provides the senior managers of NPM with more confidence to launch the innovative service to gain the first-mover advantage. Museums and other cultural institutions worldwide emphasizing the exposure and propagation of their cultural assets may learn from this finding about the ICT predicament and the processes to resolve the decision-making dilemma regarding the ICT predicament.

Zickuhr and Madden (2012) stated the fact that the growth trend of Internet usage had progressed steadily over all generations each year. The iPalace channel welcomes any visitors including web browsers, who are interested in Chinese heritage, and enjoy watching cultural videos. When other generations try to get access to the new ICT-enabled service like the iPalace channel, they may encounter the reliability problem; especially the older generation who usually has an inferior end device and limited experience with Internet usage. This reliability problem may cause for sure more frustration. Therefore, the implications regarding the ICT predicament and the decision dilemma are also useful for the RIIS targeting the older generation in the near future.

From the government's perspective, this study advises that the ICT predicament should not be overlooked if the government tries to inspire the RIISs for improving the country's economic competition worldwide. The government should provide more research aids to conduct pilot experiments to further investigate what affects the ICT infrastructure's inferiority and how this inferiority can/will affect the ICT predicament regarding the utilization of RIIS such as iPalace channel.

With the current ICT advancements, any ICT-enabled service originated from a country can be easily accessed globally via Internet. When taking the complementary ICT systems in many countries into consideration, the ICT predicament regarding some RIIS can be more difficult to be coped with than any organization may expect/predict.

Our research has the following limitations. In this study, we do not directly estimate the inferiority of ICT infrastructure and neither do we consider the antecedents of ICT infrastructure inferiority. Therefore, one of future works may be to further investigate what affects the ICT infrastructure inferiority and how the inferiority affects the ICT predicament.

Furthermore, the proposed processes entail a proper counterpart to do the comparison and perform an adequate measurement for the new service. For a specific RIIS, however, it may be difficult to identify the proper counterpart or to derive a proper measurement. In addition, the ICT predicament is also related to the RIIS itself. In this study, we do not explore how radical an innovation can cause the ICT predicament. A future work is to study the attributes of an innovation and how they would affect ICT predicament.

This study does not claim that the ICT predicament is the only challenge regarding the implementation of RIIS. For instance, the NPM does encounter with other challenges (managerial, inter-organization, and social) before launching the iPalace channel. Another future work is to comprehensively explore the challenges of RIIS.

**Appendix**

**Table A1**

The literature review of the factor of service quality in E-Commerce.

| | Groups | | | | | | |
|---|---|---|---|---|---|---|---|
| | Content | Usability | Interface | Reliability | Communication | Security | Fulfillment |
| Liu et al., (2000) | Information quality | System use | Playfulness | System design quality | | | |
| Kaynama & Black (2000) | Navigation | | Design and presentation | | Responsiveness, Personalization and customization | | Content and purpose, Background |
| Yoo & Donthu (2001) | | Ease of use | Aesthetic design | Processing speed | | Security | |
| Loiacono et al. (2002) | Ease of understanding | Intuitive operations | Visual appeal, Innovativeness, Emotional appeal, Consistent image | Response time | Tailored Communications | Trust | Informational fit-to-task, On-line completeness |
| Vidgen and Barnes (2002) | Information | Usability | Design | | Empathy | Trust | |
| Santos (2003) | Content, Structure & layout | Ease of use | Appearance | Reliability, Efficiency | Communication, Support | Security | |
| Parasuraman et al., (2005) | | Efficiency | | System availability | Responsiveness, Contact | Privacy | Fulfillment |

**Table A2**

The references of each measurement item.

| Question | References |
| --- | --- |
| Q1 | None |
| Q2 | None |
| Q3 | Loiacono et al., 2002; Vidgen & Barnes, 2002; Parasuraman et al., 2005 |
| Q4 | Liu et al., 2000; Yoo & Donthu, 2001; Vidgen & Barnes, 2002; Parasuraman et al., 2005 |
| Q5 | Yoo & Donthu, 2001; Loiacono et al., 2002 |
| Q6 | Liu et al., 2000; Loiacono et al., 2002 |
| Q7 | Liu et al., 2000; Loiacono et al., 2002; Vidgen & Barnes, 2002 |
| Q8 | Loiacono et al., 2002 |
| Q9 | Loiacono et al., 2002 |
| Q10 | Liu et al., 2000; Yoo & Donthu, 2001; Loiacono et al., 2002; Parasuraman et al., 2005; Collier & Bienstock, 2006 |
| Q11 | Liu et al., 2000; Yoo & Donthu, 2001; Loiacono et al., 2002; Parasuraman et al., 2005; Collier & Bienstock ,2006 |
| Q12 | Goulding, 2000; Rentschler & Gilmore, 2002; Harrison & Shaw, 2004; Saleh, 2005 |
| Q13 | Goulding, 2000; Rentschler & Gilmore, 2002; Harrison & Shaw, 2004; Saleh, 2005 |
| Q14 | Goulding, 2000; Rentschler & Gilmore, 2002; Harrison & Shaw, 2004; Saleh, 2005 |
| Q15 | Goulding, 2000; Rentschler & Gilmore, 2002; Harrison & Shaw, 2004; Saleh, 2005 |
| Q16 | Rentschler & Gilmore, 2002 |
| Q17 | Goulding, 2000; Rentschler & Gilmore, 2002; Harrison & Shaw, 2004 |
| Q18 | Goulding, 2000; Rentschler & Gilmore, 2002; Harrison & Shaw, 2004 |

| Q19 | McKinney et al., 2002 |
| Q20 | McKinney et al., 2002 |
| Q21 | Venkatesh et al., 2003 |
| Q22 | Venkatesh et al., 2003 |

**Table A3**

The survey instrument and the questions for text-based version

| Factor | Sub-factor | | Question |
|---|---|---|---|
| Content | Information quality | Q1 | The presented information on the website is trusted. |
| | | Q2 | The presented information on the website is easy to understand. |
| | Tele-presence | Q3 | Browsing the site, I am personally on the scene. |
| | | Q4 | Browsing the site, I am gazing steadily at the screen. |
| | Media quality | Q5 | The pictures are clear. |
| | | Q6 | The text is clear. |
| Usability | Intuitive operation | Q7 | The browsing is simple and intuitive. |
| | | Q8 | The operation is easy to learn. |
| Interface | Innovativeness | Q9 | The text-based interface is innovative. |
| | Visual appeal | Q10 | I do always focus on the content. |
| | | Q11 | The interface appearance is pleasing to both the eye and the mind. |
| | Consistent image | Q12 | There is a consistency of style in the webpage design. |
| | | Q13 | The design of the interface matches the image of the National Palace Museum. |

| Factor | Sub-factor | | Question |
|---|---|---|---|
| System Reliability | Accessibility | Q14 | I can access the web site anytime and anywhere. |
| | | Q15 | I can access the web site conveniently. |
| | Response time | Q16 | The picture and text load quickly. |
| | | Q17 | The operation of interface responds quickly. |
| Fulfillment | Education | Q18 | The website makes me have a better understanding of ancient Chinese artifacts. |
| | | Q19 | The website makes me feel more enlightened than before. |
| | | Q20 | The website makes me feel more creative than before. |
| | | Q21 | The website makes me feel more interested in the Chinese culture than before. |
| | Entertainment | Q22 | The website entertains me. |
| | Propagation | Q23 | Browsing the website is a substitute of visiting the National Palace Museum. |
| | | Q24 | Browsing the website make me feel more impressed about the National Palace Museum than before. |
| Customer Satisfaction | Satisfaction | Q25 | I am satisfied with the browsing experience of website. |
| Behavioral Intention | Recommend | Q26 | I would recommend the website to others. |
| | Reuse | Q27 | I intend to revisit the website in the near future. |
| | | Q28 | I will revisit the website in the future. |

**Table A4**

The survey instrument and the questions for the iPalace channel.

| Factor | Sub-factor | | Question |
|---|---|---|---|
| Content | Information quality | Q1 | The presented information on the website is trusted. |
| | | Q2 | The presented information on the website is easy to understand. |
| | Tele-presence | Q3 | Browsing the site, I am personally on the scene. |
| | | Q4 | Browsing the site, I am gazing steadily at the screen. |
| | Media quality | Q5 | The image of video is clear. |
| | | Q6 | The video's narration is clear. |
| Usability | Intuitive operation | Q7 | The browsing is simple and intuitive. |
| | | Q8 | The operation is easy to learn. |
| | Innovativeness | Q9 | The video-based interface is innovative. |
| | Visual appeal | Q10 | I do always focus on the content. |
| Interface | | Q11 | The interface appearance is pleasing to both the eye and the mind. |
| | Consistent image | Q12 | There is a consistency of style in the webpage design. |
| | | Q13 | The design of the interface matches the image of the National Palace Museum. |
| | Accessibility | Q14 | I can access the web site anytime and anywhere. |
| System | | Q15 | I can access the web site conveniently. |
| Reliability | Response time | Q16 | The video loads quickly. |
| | | Q17 | The operation of interface responds quickly. |
| | | Q18 | The website makes me have a better understanding of ancient Chinese artifacts. |
| | Education | Q19 | The website makes me feel more enlightened than before. |
| Fulfillment | | Q20 | The website makes me feel more creative than before. |
| | | Q21 | The website makes me feel more interested in the Chinese culture than before. |
| | Entertainment | Q22 | The website entertains me. |
| | Propagation | Q23 | Browsing the website is a substitute of visiting the National Palace Museum. |
| Customer Satisfaction | Satisfaction | Q24 | Browsing the website make me feel more impressed about the National Palace Museum than before. |
| | | Q25 | I am satisfied with the browsing experience of website. |
| Behavioral Intention | Recommend | Q26 | I would recommend the website to others. |
| | Reuse | Q27 | I intend to revisit the website in the near future. |
| | | Q28 | I will revisit the website in the future. |